# Multiferroicity and magneto-electric effect in $Gd_2BaNiO_5$

S. Chowki[1], Tathamay Basu[2], K. Singh[2,3], N. Mohapatra[1,*] and E.V. Sampathkumaran[2]

[1]IndianInstitute of Technology Bhubaneswar, Bhubaneswar, Odisha- 171013, India

[2]Tata Institute of Fundamental research, Homi Bhabha Road, Colaba, Mumbai- 400005, India

[3]UGC-DAE Consortium for Scientific Research, Indore Centre, University Campus, Khandwa Road, Indore - 452001, India

## Abstract

We report the observation of electric polarization in the magnetically ordered state of the Haldane chain compound, $Gd_2BaNiO_5$, with strongly correlated magnetic and dielectric properties. The results of dc magnetic susceptibility and heat capacity measurements indicate two magnetic transitions, one corresponding to the anti-ferromagnetic order at $T_N$ ~55 K and the other to spin-reorientation transition at $T_{SR}$ ~24 K. The dielectric permittivity($\epsilon_r^{'}$) and loss (tanδ) also exhibit anomalies in the vicinity of $T_{SR}$ and $T_N$ respectively. Below the spin-reorientation transition, concurrently magnetic-field-induced spin-flop and the meta-electric transitions are observed at a critical magnetic field in isothermal magnetization and magneto-dielectric results respectively. Another interesting finding is that $\Delta \epsilon_r^{'}$ $(=(\epsilon_r^{'}(H)-\epsilon_r^{'}(0))/\epsilon_r^{'}(0))$ changes its sign at the critical magnetic field. The origin of the observed magneto-electric effect is discussed on the basis of spin–phonon coupling.

[*]Corresponding author: niharika@iitbbs.ac.in

# Introduction

The magneto-electric (ME) multiferroics, in which magnetic and electric long range order coexist with a strong coupling between their order parameters, have attracted enormous attention recently owing to their potential application in spintronics and multi-state memory devices [1,2]. A lot of effort has been invested in the search for single phase materials exhibiting ME multiferroic behavior since the observation of giant magneto-electric effect in type II multiferroics such as $TbMnO_3$, and $HoMnO_3$ etc [3, 4]. Furthermore, the ME coupling gives rise to novel physical effects such as large magneto-dielectric (MD) effect due to strong spin–phonon coupling [5,6]. However, as the energy scale for controlling the magnetic and electric properties are different, the coupling between dielectric and magnetic properties is possible in a material in which a low frequency infrared-allowed optical phonon mode exists [7]. A recent study of the lattice and magnetic dynamics in $Gd_2BaNiO_5$ through infrared spectroscopy has shown the existence of low frequency continuum and anomalies in optical conductivity near the anti-ferromagnetic (AFM) ordering temperature ($T_N$ ~55 K) suggesting the presence of spin–phonon coupling [8]. Therefore strong ME coupling is expected in this compound. In the present work, we have investigated the electric properties of $Gd_2BaNiO_5$ through pyroelectric current and magneto-dielectric measurements.

The rare earth nickeletes, $R_2BaNiO_5$ crystallizing in *Immm*–type orthorhombic structure, exhibit intriguing magnetic properties with cooperative ordering of quantum and classical spin sublattices [9]. The crystal structure consists of 1-D chains of corner sharing $NiO_6$ octahedra running along crystallographic *a*–axis (see inset of Fig.1). Isolated chains of $NiO_6$ octahedra are magnetically interconnected via $R^{3+}$ ions. The long range AFM ordering sets in $R_2BaNiO_5$ due to the rare-earth sublattice via R-O-Ni-O-R superexchange path. Although the exchange interaction between R ions is mediated by Ni, Haldane-gap excitations were revealed by neutron diffraction experiments in the magnetically ordered state as well [10]. Therefore the resultant magnetic behavior of $R_2BaNiO_5$ can be explained as the coexistence of the gapped spin-singlet state with AFM long range ordered state. Another interesting property of this family is multiferroicity and MD effect, which has been reported recently for some of the members (R = Ho, Dy & Er) [11-13], while the other members remain unexplored for these phenomena. Therefore, we consider another member of this Haldane spin-chain system, $Gd_2BaNiO_5$, as a potential candidate for the study of magneto-electric behavior.

## Experimental details

Polycrystalline sample of $Gd_2BaNiO_5$ was synthesized using the conventional solid state reaction route as described in literature [14]. The phase-purity of the sample was examined by X-ray powder diffraction technique using PANalytical diffractometer with Cu-K$\alpha$ radiation. The dc magnetic susceptibility ($\chi$) and Isothermal magnetization (M) measurements were performed with Quantum Design superconducting quantum interference device (SQUID) magnetometer in the temperature (T) range 1.8–300 K. The heat-capacity (C) and electrical resistance measurements were carried out with a Quantum Design physical properties measurement system (PPMS). A home-made sample holder was integrated with the commercial PPMS for the measurement of capacitance (as described in Ref. 12). The capacitance and dielectric loss (tan$\delta$) measurements as a function of temperature and magnetic field were made in the frequency range 5–100 kHz using Agilent E-4908A LCR-meter. Isothermal magneto-dielectric behavior was measured with 1V ac bias at 100 kHz by ramping magnetic field at the rate of 100 Oe/sec. The temperature variation of dielectric constant was recorded while warming at 1K/min. Remnant polarization (*P*) as a function of *T* was measured with Keithley 6517A electrometer.

## Results

The room temperature XRD pattern of $Gd_2BaNiO_5$ was analyzed by Rietveld refinement method using the FullProf Suite program [15]. As shown in Fig. 1, all the reflections could be indexed to the orthorhombic crystal structure with *Immm* symmetry. The refined lattice parameters (*a* = 3.7885(2) Å, *b* = 5.7554(3) Å and *c* = 11.3316(6) Å) are in good agreement with the previously reported results [16].

The results of magnetization measurements obtained in presence of 100 Oe and 5 kOe magnetic fields are shown in Fig. 2(a) in the T range 2–80 K. Curie-Weiss behavior of χ(T) is observed above 100 K (see inset of Fig. 2(a)) with an effective moment ($\mu_{eff}$) of 11 $\mu_B$/f.u., which is close to the theoretical value of trivalent Gd. The negligible contribution of Ni sublattice to the effective moment may be due to the low-dimensional magnetic interaction. These results are in good agreement with those reported in literature [17]. There is a sharp anomaly in χ(T) around ~24 K attributable to spin reorientation transition (SRT) which is followed by an increase on further lowering temperature for both 100 Oe and 5 kOe applied field [17,18]. Earlier reports based on the microscopic studies such as optical [8] and Mössbauer spectroscopy [19] of $Gd_2BaNiO_5$ have shown the presence of long-range AFM ordering below ~55 K ($T_N$). Although we did not observe any feature in χ(T), the derivative plot d(χ(T)T)/dT as a function of temperature (shown in the same figure) exhibits a change of slope near $T_N$ indicating the onset of magnetic order. It is to be noted that a distinct SRT in χ(T) is observed only for $Gd_2BaNiO_5$, while all other members of $R_2BaNiO_5$ family exhibit a broad maximum attributable to the low dimensional magnetic interactions [17].

The plot of C/T as a function of temperature is shown in Fig. 2(b). In the absence of magnetic field, it exhibits a sharp peak ~24 K ($T_{SR}$) and a small anomaly near 55 K ($T_N$). As shown previously in the study of single crystal sample [18], a part of magnetic entropy is released presumably due to short range ordering above $T_N$ as a consequence of which a weak anomaly only is seen around $T_N$. In addition to these major transitions, the C(T)/T data reveals another broad peak around 6 K. The origin of this broad peak is not clear to us at this moment although it has been attributed to Schottky type excitations of the ground state sublevels of $Gd^{3+}$ ion [18]. With the application of magnetic field, the anomaly at 24 K and 6 K gets suppressed while the peak at 55 K remains unaffected (Fig. 2 (b)).

In the inset of Fig. 2(b), we have shown the electrical resistivity measured as a function of temperature by two probe method. The room temperature resistivity of $Gd_2BaNiO_5$ was found to be ~40 MΩ-cm which keeps on increasing with decreasing temperature. We could measure resistivity down to 80 K only and below this temperature the resistance of the sample exceeded beyond the capability of the experimental setup. It is also to be noted that there is no noticeable magneto-resistance in this compound.

Fig. 3(a) and (b) depicts the temperature dependence of dielectric constant ($\epsilon'_r$) and dissipation factor (tanδ) in zero applied magnetic fields at selected frequencies. As the magnetic transitions occur in the low T region (< 80 K), we discuss the magneto-dielectric correlation in this T range only. By lowering the temperature below 80 K, $\epsilon'_r(T)$ exhibits a smooth decrease through the AFM transition and a minimum near SRT (~24 K) which is followed by an increase. The minimum in $\epsilon'_r(T)$ is found near 24 K ($T_{min}$) for 100 kHz frequency and it shifts towards lower T for lower frequencies ($T_{min}$= 15K for 5 kHz, see inset of Fig. 3(a)). On the other hand, a well-defined shoulder in tanδ is observed around $T_N$ which becomes prominent with increasing frequency, shifting towards higher T. Additionally, a broad peak is observed around 6 K which is found to be frequency independent. It should be noted that C(T)/T also exhibits an anomaly at the same temperature. As shown in the inset of Fig. 3(b), this low T peak gets suppressed in the presence of high magnetic field.

To get further insight into the dielectric anomalies and its correlation with lattice dynamics, we analyzed $\epsilon'_r(T)$ vs T data on the light of modified Barrett's theory [20] predicted for dielectrics. The theory provides a link between temperature dependence of the dielectric constant with the excitation of low-lying optical phonons by the relation, $\epsilon'_r(T) = \epsilon'_r(0) + \frac{A}{\left[\exp\left(\frac{\hbar\omega_0}{k_B T}\right)-1\right]}$, where $A$ is the coupling constant and $\omega_0$ is the mean frequency of the final states in the lowest lying optical phonon branch. A good agreement of $\epsilon'_r(T)$ with the Barrett equation is realized in the temperature range 60–80 K i.e. just above $T_N$ (see inset of Fig. 3(b)). The values of $A$ and $\omega_0$ obtained from the fit are 141.2 and 103 cm$^{-1}$ respectively. Recent infrared spectroscopy study on single crystal of $Gd_2BaNiO_5$ shows an optical phonon mode at ~80 cm$^{-1}$ along the chain direction (E∥a), while the optical conductivity at this frequency shows pronounced T-dependence with a kink near $T_N$ [8]. The small difference in the frequency value obtained from the $\epsilon'_r(T)$ data may be due to the use of polycrystalline sample in the present study. It is to be noted that the fit of Barrett equation with $\epsilon'_r(T)$ deviates just above $T_N$ suggesting a shifting of the frequency around this temperature similar to the behavior observed in TbFe$_3$(BO$_3$)$_4$ [21].

Pyroelectric current measurements were performed to explore the presence of remnant polarization in the magnetically ordered state. The sample was cooled down to 7 K in a poling field of 370 kV/m and then shorted after switching off the electric field to discharge any stray current. Pyroelectric current was recorded while warming the specimen at a rate of 2 K/min in zero applied fields. Fig. 4 shows the temperature dependence of polarization which was obtained by integrating the pyroelectric current with respect to time. Interestingly the polarization starts developing just above $T_N$ which increases with lowering temperature and finally saturates below ~38 K. The sign reversal of polarization by reversing the direction of poling electric field confirms ferroelectric state of the sample below $T_N$. The observed magnitude of polarization (P ≈ 8 μC/m$^2$), although small for a single domain ferroelectric material, is comparable to that observed in Dy$_2$BaNiO$_5$ [12] and other multiferroics in the spinel group such as CoV$_2$O$_4$ [22] and CoCr$_2$O$_4$ [23]. It is also to be noted that a peak in the pyroelectric current is observed around 50 K both at positive and negative poling fields indicating a transition from para-to-ferroelectric state near $T_N$ as shown in the inset of Fig.4.

To further elucidate the relationship between magnetic and dielectric characteristics, we have shown the results of *M(H)* and $\Delta\epsilon'_r$ (=($\epsilon'_r(H)-\epsilon'_r(0))/\epsilon'_r(0)$) measurements at selected temperatures in Fig.5(a) and (b) respectively. In the T range $T_{SRT}> T > T_N$, say at 50 K, $\Delta\epsilon'_r$ is found to be small and positive. *M(H)* is found to vary linearly with *H* in this T range. The non-linearity of *M(H)* develops just below $T_{SR}$ which becomes pronounced by further lowering T. At 2 K, a change of slope near a critical field (H$_C$) of 12 kOe indicates the magnetic-field-induced spin-flop transition. Moreover, this H$_C$ increases with increasing T. Remarkable change in the magnitude of $\Delta\epsilon'_r$ is also realized below $T_{SR}$. The maximum value of $\Delta\epsilon'_r$ is found to be -4% at 5 K for an applied magnetic field of 140 kOe. The most striking feature is the observation of magnetic-field-induced effects in both *M(H)* and $\Delta\epsilon'_r$ concurrently occurring at the same field. While *M(H)* data show only a change of slope around $H_C$, a pronounced peak is noticed in $\Delta\epsilon'_r$ at $H_C$. Such a peak in $\epsilon'_r(H)$ has been ascribed to a meta-electric transition, an electric analogue to meta-magnetic transition, induced by magnetic field as previously reported in BiMn$_2$O$_5$ [24]. The critical-field inducing meta-electric transition is also found to increase with increasing temperature. This increase in critical magnetic-field in dielectric results is consistent with *M(H)* results on single crystal sample as reported by Popova et al. [18] and also with our *M(H)* results, showing the strong magneto-dielectric effect. It is also to be noted that $\Delta\epsilon'_r$ shows a crossover from negative to positive sign following the meta-electric transition.

The above results suggest multiferroic behavior of Gd$_2$BaNiO$_5$ with magneto-dielectric effect. To understand the spin-lattice-charge correlations we have discussed our results on the basis of the temperature and magnetic field dependent phase diagram established by Popova et al [18] by the investigations of thermodynamic and magnetic measurements performed on a single crystal. In zero field, Gd and Ni moments lie in the *ac* plane in the T range $T_{SR} < T < T_N$ and orient in the bc plane below $T_{SR}$. Further, by the application of magnetic field, the moments flop to the ac plane below $T_{SR}$ and to the bc plane in the T range $T_{SRT} < T < T_N$. In the present study, we have shown dielectric permittivity exhibiting sharp meta-electric transition at the same magnetic-field that induces spin-flop which suggests the sensitive nature of $\epsilon'_r$ to magnetic interactions. In a recent study of the dielectric properties of Dy$_2$BaNiO$_5$, another member of this family, we have shown that the dielectric anomalies are concomitant with magnetization behavior which depends on different orientations of Ni$^{2+}$ and Dy$^{3+}$ magnetic moments [12, 25]. Now relating the orientations of Ni$^{2+}$ and Gd$^{3+}$ moments with $\Delta\epsilon'_r$ values, it is found that $\Delta\epsilon'_r$ is positive when Ni$^{2+}$ and Gd$^{3+}$ moments are in the *bc* plane and negative while both these moments orient in the *ac* plane.

In addition, the Raman and infrared spectroscopy studies also establish the strong correlation of magnetic excitations with lattice dynamics. The frequency and half width of the observed phonon modes exhibit pronounced temperature dependence peaking at $T_N$ [8]. It has also been shown that the magnetic absorption, which is a predominantly two magnon optical absorption, has been attributed to the spin-phonon interaction. Therefore, our results along with the available magnetic phase diagram and optical studies demonstrate possible correlation of magnetic and electric properties via phonons in the Haldane chain system Gd$_2$BaNiO$_5$.

## Conclusions

We have investigated the magnetic and electric properties of one of the Haldane chain antiferromagnet Gd$_2$BaNiO$_5$. Anomalies are observed in magnetic susceptibility, dielectric permittivity, tanδ and pyroelectric current at the onset of long range AFM and spin reorientation ordering. The development of electric polarization near T$_N$ implies a correlated behavior of magnetism and ferroelectricity, categorizing this compound in type-II multiferroic materials. Furthermore, magnetic-field-induced spin flop transition is found to be associated with the novel meta-electric transition. Crossover of positive to negative values of $\Delta\epsilon_r^{'}$ may be attributed to the change of orientation of Ni$^{2+}$ and Gd$^{3+}$ moments from *bc* plane to *ac* plane. These findings suggest a correlated behavior of spin-charge-lattice degrees of freedom and magneto-electric coupling in this compound.

## Acknowledgement

The authors thank Mr. Kartik K Iyer, Tata Institute of Fundamental Research, Mumbai for his help in the experimental work.

**Figure Captions:**

**Fig. 1**: (Color online) The room temperature XRD pattern of $Gd_2BaNiO_5$. Circles represent the experimental data while solid curve is the best fit from the reitveld refinement using Fullprof. The position of Bragg reflections are marked by vertical lines. Inset: crystal structure drawing of $Gd_2BaNiO_5$ in which chains of $NiO_6$ octahedra are well separated by $R^{3+}$ and $Ba^{2+}$ ions.

**Fig. 2**: (Color online) (a) Temperature dependence of magnetic susceptibility ($\chi(T)$) of $Gd_2BaNiO_5$ (left panel) and derivate of $\chi(T)T$ with respect to T (right panel). Inset shows inverse susceptibility as a function of T with the line showing Curie-Weiss fit in the paramagnetic state. (b) Heat capacity of $Gd_2BaNiO_5$ as a function of temperature in presence of various external magnetic fields. Inset shows the variation of electrical resistivity with T.

**Fig. 3:** (Color online) (a) The relative dielectric permittivity ($\epsilon_r'$) versus temperature measured at different frequencies in zero field. Inset shows the anomaly in $\epsilon_r'$ near $T_{SRT}$. (b) Temperature dependence of the loss (tan$\delta$) measured at different frequencies in the absence of external magnetic-field. Inset (I) shows the effect of magnetic field on the low temperature anomaly in $\epsilon_r'$ at 100 kHz. Inset (II) shows the fit to the Barrett's equation as described in the text.

**Fig. 4**: (Color online) Temperature dependence of remnant polarization in the absence of external magnetic field. Inset shows the pyroelectric current as a function of T.

**Fig. 5:** (Color online) Magnetic field dependence of (a) magnetization and (b) percentage change of dielectric permittivity, $\Delta\epsilon_r'$ $(=(\epsilon_r'(H)-\epsilon_r'(0))/\epsilon_r'(0))$ measured at selected temperatures at 100 kHz.

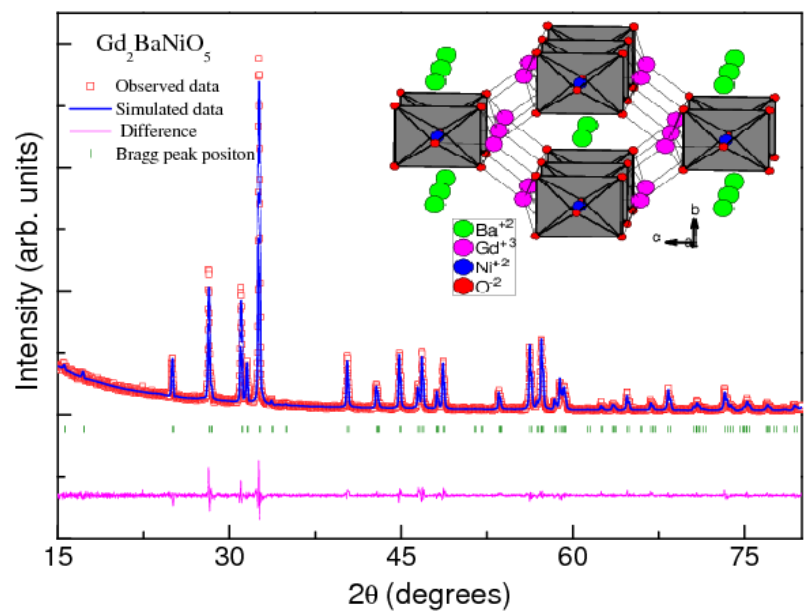

**Fig.1**

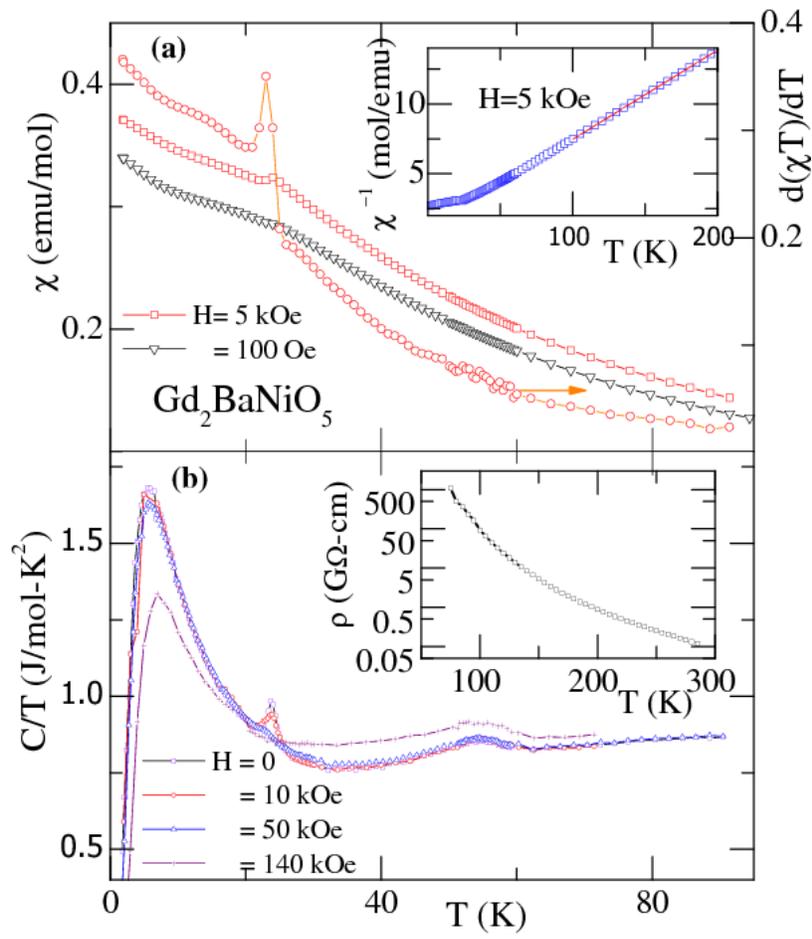

**Fig. 2**

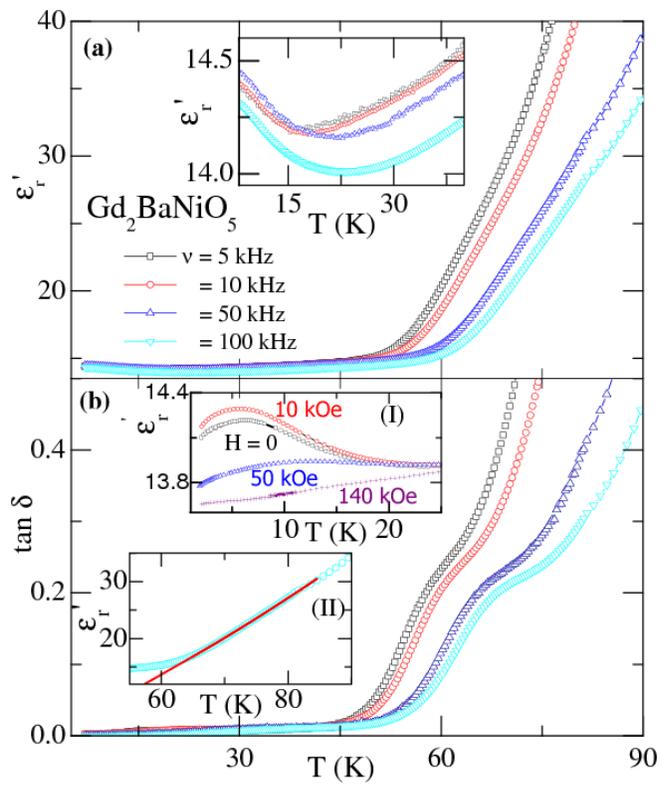

**Fig. 3**

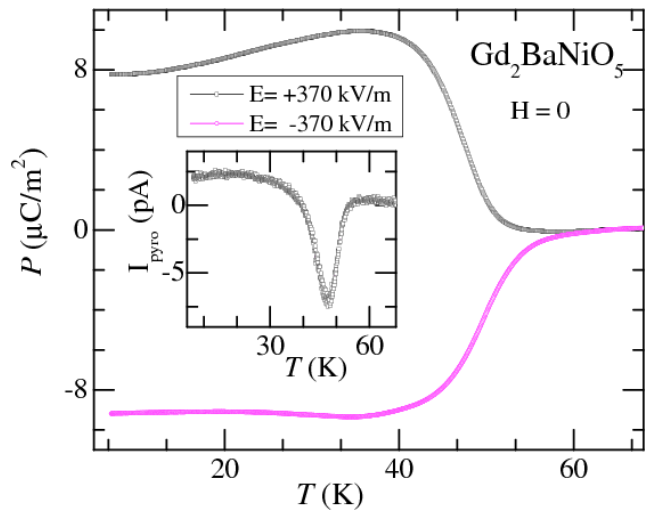

**Fig.4**

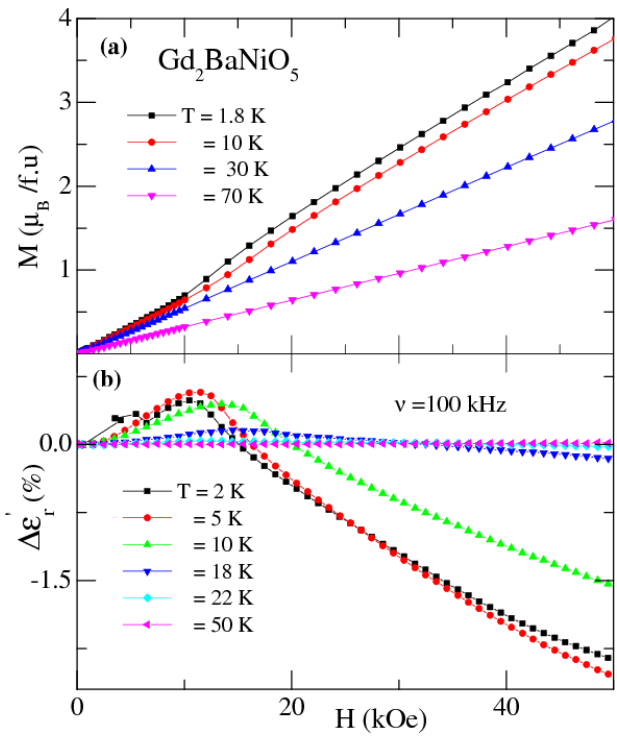

**Fig.5**